# Integrating Generative AI in Hackathons: Opportunities, Challenges, and Educational Implications

**Ramteja Sajja**[1,3] (ramteja-sajja@uiowa.edu)
**Carlos Erazo Ramirez**[2,3] (carlosvalentin-erazoramirez@uiowa.edu)
**Zhouyayan Li**[2,3] (zhouyayan-li@uiowa.edu)
**Bekir Z. Demiray**[3,4] (bekirzahit-demiray@uiowa.edu)
**Yusuf Sermet**[3] (msermet@uiowa.edu)
**Ibrahim Demir**[1,2,3,4] (ibrahim-demir@uiowa.edu)

[1] Department of Electrical Computer Engineering, University of Iowa
[2] Department of Civil and Environmental Engineering, University of Iowa
[3] IIHR Hydroscience and Engineering, University of Iowa
[4] Informatics, University of Iowa

**Abstract**
Hackathons have emerged as pivotal platforms in the software industry, driving both innovation and skill development for organizations and students alike. These events enable companies to quickly prototype new ideas while offering students practical, hands-on learning experiences. Over time, hackathons have transitioned from purely competitive events to valuable educational tools, integrating theory with real-world problem-solving through collaboration between academia and industry. The infusion of artificial intelligence (AI) and machine learning is now reshaping hackathons, providing enhanced learning opportunities while also introducing ethical challenges. This study explores the influence of generative AI on students' technological choices, focusing on a case study from the 2023 University of Iowa Hackathon. The findings offer insights into AI's role in these events, its educational impact, and propose strategies for integrating such technologies in future hackathons, ensuring a balance between innovation, ethics, and educational value.

## 1. Introduction

In the ever-evolving software engineering and development landscape, hackathons have risen to prominence, serving as crucial platforms for companies to swiftly convert ideas into tangible software prototypes. This agility grants them a competitive advantage in the unpredictable business world (Komssi et al., 2014). In addition, hackathons promote research, collaboration, and education between students and interested parties like companies and educational organizations (Briscoe, 2014). Students and researchers use hackathon for developing new software libraries and frameworks (Ramirez et al., 2022; Erazo et al., 2023a). Moreover, the inclusion of these entities promotes fast-track solutions in a restricted time frame (Cobham et. al., 2017).

Over the past two decades, hackathons and coding camps have also had an impact in the educational domain (Porras et al., 2005). They represent a nexus between theory and hands-on problem-solving. Their integration into computer science and software engineering syllabi has been diverse, yet the underlying aim remains to focus on increasing specific skill sets while resonating with curriculum objectives (Porras et al., 2019). The main objective of these events is to align educational skills within a larger collaborative learning context. Far beyond being solely a competitive event, as observed in platforms like annual university hackathons, these are safe and controlled environments for peer connectivity and learning (Nandi & Mandernach, 2016). The confluence of expertise from both the academic and industrial sectors amplifies learning, rendering invaluable mentorship and hands-on insights (Medina Angarita & Nolte, 2020).

Hackathons have emerged as a pivotal platform for both companies and students. For companies, they are tools to test products, bolster their brand, and scout for talent (Pe-Than et. al., 2018). Companies and academia provide scientific challenges and benchmark datasets (Sit et al., 2021; Demir et al., 2022) to bridge domain and data science collaboration (Ebert-Uphoff et al., 2017). Simultaneously, students harness these events as unparalleled learning avenues. They delve into advanced software tools and libraries (Erazo et al., 2023b), hone their programming skills beyond classroom confines, and grapple with real-world tasks in environments marked by ambiguity (Lara et al., 2015). Collaborating firms frequently present participants with genuine business conundrums accompanied by real-world data. These challenges span a diverse array of sectors, from energy and climate change to agriculture and insurance, offering students a broadened perspective. Such experiences are instrumental in sharpening both their technical and interpersonal proficiencies and provide them with valuable portfolio assets to propel their careers forward (Lara & Lockwood, 2016).

Recently, AI-powered tools have emerged as disruptive technologies within educational settings, particularly in computing education. Tools like ChatGPT and GitHub Copilot, released between 2022–2023, have gained significant attention from computing educators. These tools can generate correct solutions to programming assignments and explain code content, raising concerns among educators about adapting courses to ensure effective student learning (Lau & Guo, 2023). Some educators aim to discourage AI-assisted cheating, while others advocate integrating AI into curricula to prepare students for the evolving job market (Lau & Guo, 2023). Generative models are also being used to generate programming exercises and code explanations, with varying levels of acceptance among students and instructors. As Zastudil et al. (2023) observed, while generative AI tools offer significant potential in computing education, concerns remain about how best to integrate them to support student learning goals. The rapid adoption of AI in classrooms calls for an exploration of stakeholder preferences to mitigate potential harms and ensure these tools are used for good (Zastudil et al., 2023).

In parallel, the proliferation of generative AI tools like ChatGPT offers students new help-seeking resources that are always available. However, students continue to rely on traditional help resources depending on the task, and AI help-seeking skills are still developing, benefiting those adept at harnessing these tools (Hou et al., 2024). The introduction of AI tools into education represents both a challenge and an opportunity for educators, who must adapt to these emerging resources to enhance students' learning experiences.

The transformative impact of AI is already being witnessed in computing education through the deployment of AI-based tools in courses such as Harvard University's CS50 (Liu et. al., 2024). These tools provide continuous, customized support for students, simulating a near 1:1 teacher-to-student ratio and augmenting traditional pedagogical approaches. As Liu et al. (2024) reported, such tools are well-received by students and improve learning outcomes, demonstrating the potential for AI to enhance the teaching experience in computing education. Similarly, comprehensive explorations of AI in computing education, such as the work by Prather et al. (2023), highlight the rapid improvements in AI tools and their profound potential to reshape educational practices.

As the integration of AI continues to grow, its utility in hackathon environments is also becoming more apparent. The introduction of large language models (LLMs) like GPT into hackathon settings provides students with advanced technological capabilities, influencing their project development choices and collaboration methods. Sajja et al. (2024) demonstrated the potential of AI in environmental science education, showcasing how tailored educational AI assistants can improve student engagement and learning outcomes through seamless integration with learning management systems (LMS). Similarly, hackathons can benefit from AI tools that offer personalized support and enhance the learning experience in complex and data-intensive fields.

Diving deeper into the educational dimension of hackathons, studies have highlighted their transformative learning potential. La Place et al. (2017) study highlighted how students quickly adapt and learn new skills on the fly. Warner et al. (2017) found that many students join hackathons to boost their skills and for social reasons, though some feel intimidated. Gama et al. (2018) even incorporated hackathons into classroom settings, blending traditional learning with real-world challenges. Overall, hackathons provide a unique opportunity for students to grow, collaborate, and prepare for real-world job scenarios.

Yet, as hackathons evolve, they are becoming intertwined with advanced technologies, notably artificial intelligence (AI) and machine learning. Such integrations are revolutionizing learning and research methodologies. Related technologies, such as chatbots, are changing how we learn, conduct and communicate research (Sermet and Demir, 2018). However, as Kooli (2023) points out, this combination raises ethical concerns. There should be a critical balance between embracing AI in education and ensuring ethical and educational integrity. Going further into higher education, Chan (2023) suggests a need for a comprehensive AI education plan. This plan should address the complex effects of AI and involve all educational stakeholders. Additionally, the rise of Generative Pre-trained Transformers (GPT), like OpenAI's ChatGPT, brings another aspect to the discussion. As observed by Grassini (2023) on these technologies' remarkable abilities, such as generating human-like text or code, these models raise concerns about their impact on critical thinking skills and their practical use.

Bridging these observations to the hackathon setting, it's evident that the introduction of such emerging technologies can revolutionize technology learning experiences. Hackathons stand to benefit immensely from AI-driven tools. However, to harness their full potential while mitigating associated risks, it's imperative to adopt a measured and informed approach (Krishnamurthy et al., 2018). By doing so, hackathons can remain a crucial part of tech education, driving innovation while maintaining ethical and educational integrity.

This study explores the impact, utility, and potential of generative AI on hackathons, focusing on the hackathon at the 2023 University of Iowa Hackathon. It examines how AI technologies, especially large language models like GPT, influence students' technological choices and project development. The research highlights the growing integration of AI in hackathons, analyzing its effects on efficiency, learning experience, and collaboration. It also addresses the challenges and ethical concerns raised by AI usage, providing insights into the balance between innovation and educational integrity in the evolving landscape of hackathon events.

The paper is continuous with the methodology section which describes an exploratory analysis conducted via a series of research questions asked during and after the event. The results section presents the major survey results and specific insights into the use of generative language models on various topics explored during the event. The discussion section delves into the implications of these technologies on hackathon events, their significance for research and education, and offers recommendations for future events and open research directions. The study concludes with a summary of the findings.

## 2. Methodology

In the dynamic world of technology and education, the approach to gathering, analyzing, and interpreting data is paramount to producing meaningful insights. The methodology section outlines the strategies and tools we employed to examine the intersection of hackathons, education, and the evolving role of AI tools like ChatGPT. This section outlines our research process in order to guide replication and future research in this area. From the design of our survey to the rationale behind each question, this section offers a comprehensive view into our investigative journey. Through our methodology, we aim to capture the nuances and complexities of the hackathon experience in the age of AI, setting the stage for informed discussions and implications for both the tech and educational sectors.

### 2.1. Event Description for Survey Context

The 2023 edition of an annual hackathon organized by a computer science department at the 2023 University of Iowa Hackathon marks a notable event in the realm of technology and innovation. This 36-hour competition is designed for university students who have a keen interest in technological advancements and creative problem-solving. It offers a dynamic environment for participants to form teams, with up to four members, to develop projects across a variety of platforms including web, mobile, desktop, and hardware applications. The event was designed to promote creativity and technological exploration among students of all skill levels. Moreover, mentors from academic faculties and industry partners are available to offer guidance, significantly enriching the educational value of the hackathon. Beyond serving as a crucible for technological development, the event also facilitates networking opportunities with professionals in the industry, potentially bolstering the career prospects of its participants.

## 2.2. Survey Objectives and Background

With the growing use of tools like ChatGPT in the tech world, we need to think about how they might affect hackathons. Traditionally, hackathons have been about coming up with new ideas, learning, and working together. While these tools can be useful, they might also change the way we see and value hackathons. So far, those emerging tools have well-recognized boosting effects in quite a few applications, and at the same time, raised concerns in at least the following aspects as seen in Table 1.

Table 1. Impact of AI Tools on Hackathons: Considerations and Effects

| **Aspect** | **Impact of AI Tools** | **Considerations** |
| --- | --- | --- |
| Efficiency and Project Quality | Enhances coding and debugging efficiency. Facilitates idea generation leading to more sophisticated projects within event constraints. | Balancing tool-assisted development with original thinking. |
| Learning Experience | May limit problem-solving skills due to reliance on AI for solutions but may provide new insights & directions, enriching the learning process. | Ensuring a balance between tool usage and independent learning & problem-solving. |
| Collaboration | Over-reliance on AI tools could bypass team discussions and debates, impacting the collaborative spirit. | Encouraging team interactions & discussions alongside AI tool utilization. |
| Leveling the Playing Field | Bridges the knowledge gap between novices and experienced developers, allowing beginners to contribute more meaningfully. | Maintaining a fair and inclusive environment for all skill levels. |
| Evaluation and Ethics | Raises questions about the evaluation of projects and the ethical use of AI-based tools in development of solutions. | Developing clear guidelines & criteria for AI tool usage and project evaluation. |

Given these considerations and the potential implications of these tools on hackathons, this survey aims to investigate the research questions (RQ) described in Table 2.

## 2.3. Overview of Participants

The hackathon at the 2023 University of Iowa Hackathon gathered a wide array of participants, primarily students with a keen interest in technology-related disciplines. Attracting 233 attendees, this event drew young tech aficionados from various parts of the country. The majority of these participants were enrolled in undergraduate and graduate programs across a range of universities in the United States, fostering a melting pot of diverse perspectives, experiences, and skills. Key areas of study included Computer Science and Computer Engineering, focusing on both theoretical and practical aspects of computer systems and design. There was also notable participation from students in Informatics programs, addressing the intricacies and possibilities of managing significant volumes of data in today's digital era. This eclectic group not only spurred competitive spirit but also promoted the sharing and refinement of ideas, techniques, and best practices among the participants.

Table 2. Study Research Questions

| RQ | Description |
| --- | --- |
| RQ1 | How extensively are natural language processing (NLP) technologies, including ChatGPT, being integrated into hackathon projects? |
| RQ2 | For which specific aspects of project development (e.g., debugging, idea generation) is ChatGPT most frequently utilized? |
| RQ3 | Is there a perceived reduction in the importance of manual coding skills due to tools like ChatGPT and GitHub Copilot among hackathon participants? |
| RQ4 | Would AI-driven communication tools in platforms like Discord enhance collaboration and efficiency during hackathons? |

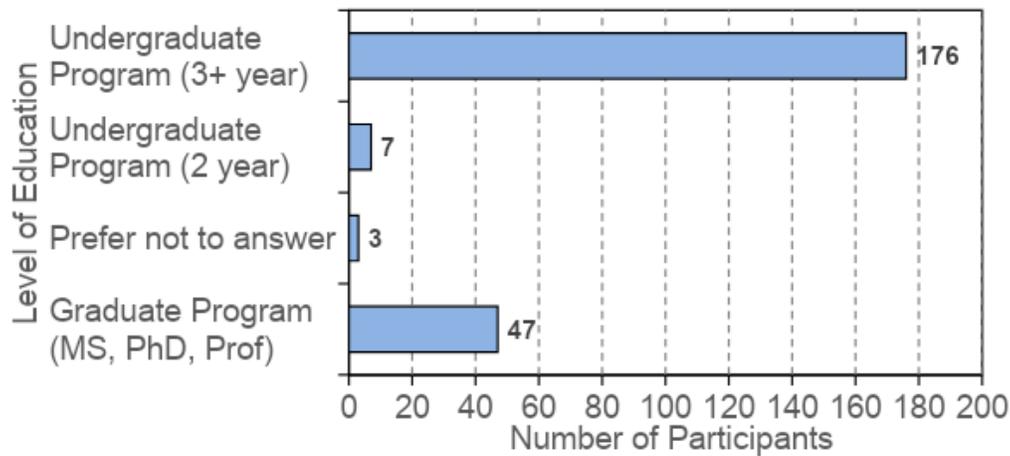

Figure 1. Educational background of participants.

Figure 1 categorizes the educational background of the hackathon at the 2023 University of Iowa Hackathon participants, indicating a majority of attendees are pursuing extended undergraduate programs, with 176 individuals enrolled in universities for three years or more. The graph also shows a relatively lower engagement from community college students, with only 7 participants from such two-year programs. Graduate-level participants, including those pursuing masters, professional, and doctoral degrees, are also a significant presence, totaling 47. A small fraction, consisting of 3 participants, opted not to specify their educational level. This data points to the broad appeal of the hackathon, suggesting it is a forum well-suited to students from diverse academic stages, from undergraduate to graduate levels, fostering a comprehensive educational experience.

Figure 2 depicts the age distribution of the hackathon at the 2023 University of Iowa Hackathon participants, highlighting a youthful demographic with the majority being 21 years old or younger. The graph shows a decline in participation for ages beyond this, with a modest presence of older participants. This pattern suggests that the event primarily attracts college-age individuals, which aligns with the typical age range of university students and recent graduates who are often active in the hackathon circuit.

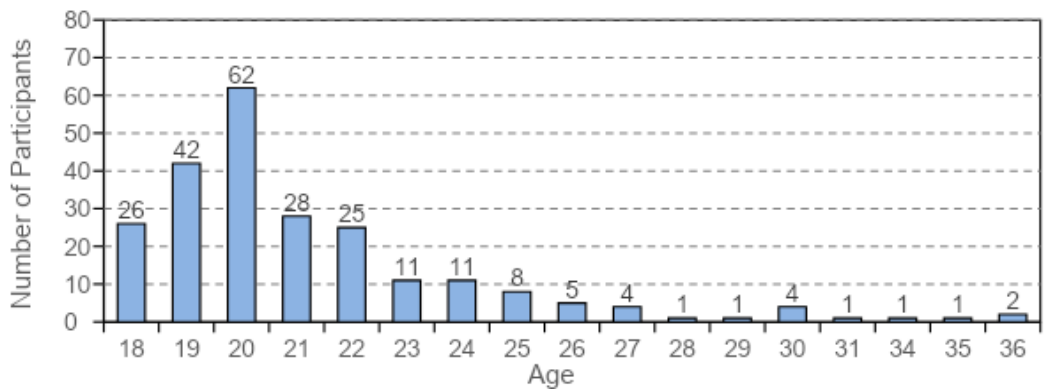

Figure 2. Age distribution of hackathon participants.

## 2.4. Overview of Survey Questions

As we navigate the evolving landscape of technology and its intertwining with education, understanding the nuanced impacts of tools like ChatGPT on hackathons becomes essential. Hackathons, traditionally known for fostering innovation, collaboration, and hands-on learning, now operate in an environment abundant with cutting-edge tools. This section delves into the key questions, as seen in Table 3, posed to the participants of the hackathon at the 2023 University of Iowa Hackathon. These questions aim to shed light on the role and perception of advanced AI tools in this context. Each query is grounded in a specific rationale, and Table 3 provides a detailed overview of these questions and the purpose behind each, serving as a roadmap to the findings and discussions that follow.

**Table 3**: Overview of Questions and the reasoning behind them

| Question | Purpose |
| --- | --- |
| What type of project did your team primarily focus on during the hackathon? | *To determine the variety of project themes and domains that participants choose, which can offer insights into current trends and interests among student developers.* |
| Which programming languages did your team primarily use for the project? | *To gauge the most popular and widely used programming languages in contemporary hackathons, indicating current educational emphasis and industry relevance.* |
| Which frameworks or libraries did your team primarily rely on for building your hackathon project? | *Understanding the frameworks and libraries in use can shed light on the complexity of the projects, the tools students are familiar with, and potential areas of focus in tech curriculums.* |
| To what extent did your team utilize ChatGPT in your project? | *To assess how integral emerging AI tools, like ChatGPT, are in the hackathon development process, suggesting potential shifts in traditional coding practices.* |
| On a scale from 1 to 5, how valuable were NLP technologies (including ChatGPT) in achieving your project goals? | *This aims to quantify the perceived value and impact of NLP tools in project outcomes, helping to determine if such technologies are seen as enhancements or crutches.* |
| Which aspects of your project did you use ChatGPT or ChatGPT API for? | *To dissect specific areas where AI tools are most beneficial or frequently employed, such as debugging or idea generation, and gauge their breadth of application.* |
| In your opinion, do coding assistance technologies like ChatGPT or GitHub Copilot reduce the need for manual coding skills during hackathons? | *This seeks to understand the perceived implications of AI tools on foundational coding skills, hinting at potential shifts in educational needs or hackathon strategies.* |
| Would integrating an AI agent into Discord for hackathon communication improve the overall experience and efficiency of communication with students? | *To assess the openness to integrating AI into communication platforms, which can offer insights into potential avenues for improving hackathon organization and collaboration.* |
| Which is more helpful for answering questions during hackathons in platforms like Discord? | *Aiming to identify preferred sources of assistance and information during hackathons, this can guide organizers in tailoring support mechanisms for participants.* |
| To what extent do you believe that technologies like ChatGPT, which can generate code, impact the traditional hackathon experience? | *This broad question seeks to capture general sentiments around the fusion of AI in hackathons, gauging whether such tools are seen as transformative, beneficial, or potentially detrimental.* |

## 3. Results

In this section, we delve into the empirical insights gathered from the survey, shedding light on participants' perceptions and experiences related to AI technologies, particularly ChatGPT, in the realm of hackathons. Drawing from a diverse pool of 151 respondents, these findings offer a comprehensive understanding of the evolving dynamic between traditional hackathon practices and the infusion of AI-driven tools. Navigate through the subsections to discern the nuanced perspectives on the utility, impact, and potential of these technologies in shaping the future of collaborative coding events.

## 3.1. Key Project Categories and Utilized Technologies

This subsection explores the main project categories that hackathon participants selected during the competition. The choices made by teams can provide a deeper understanding of the prevalent technology trends and domains. Hackathons serve as a vibrant platform for innovation, bringing together a wide array of creative and technological ideas. This year's event displayed a diverse range of project themes, reflecting the evolving interests and skills within the tech community. The data, as depicted in Figure 3, offers a comprehensive view of these trends.

Web Development emerged as the predominant choice among participants, with an impressive 49% of teams focusing on this domain. This trend can be attributed to the accessible nature of web development, the abundance of web tools and templates, and the general proficiency in web-based languages and frameworks among the participants. Following this, 17% of participants delved into projects related to Data Science or Analytics, while Artificial Intelligence projects engaged 14% of participants.

Other categories like Game Development, Mobile App Development, and Hardware or IoT (Internet of Things) garnered balanced attention, each attracting between 5% and 7% of participants. Additionally, niche areas like Browser Extensions, Cybersecurity, and other similar projects categorized as "other" accounted for 3% of the participation. This reflects the expanding scope of technology and the eagerness of participants to explore emerging fields. This distribution highlights both the dominance of web-centric projects and the wide-ranging tech interests present in modern hackathons.

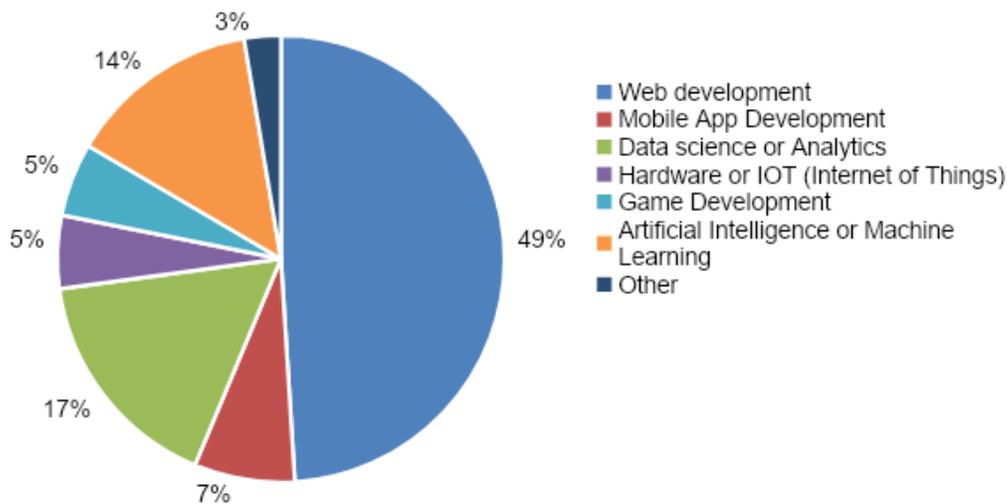

Figure 3. Distribution of project types at the hackathon.

**Programming Language Selection**

The selection of programming languages in a hackathon setting is a pivotal decision that can greatly influence the trajectory and outcome of a project. This section examines the popular programming languages among participants, focusing on those that have become integral to achieving success in hackathons. As illustrated in Figure 4, we observe the distribution of primary programming languages used by teams in their projects. JavaScript emerged as the most prevalent choice, with 104 participants opting for it. This is reflective of JavaScript's essential role in web development, aligning with the earlier observation of web development being the foremost project theme at the hackathon. The popularity of JavaScript can be attributed to its versatility in front-end development, ease of integration with various web technologies, and its expansive ecosystem of frameworks and libraries.

Closely following was Python, which was the primary choice for 93 participants, showcasing its versatility in various tech domains including web backend, data science, and artificial intelligence. The other languages shown witnessed a considerably lower usage rate: C++ was chosen by 10 participants, C# by 6 participants, and 28 participants for other programming languages not specified in the graph. It's worth noting that students often used more than one language in their projects, and each entry was calculated independently. The "Other" category encompasses languages (i.e., Java, TypeScript, R, KML, etc.) that were below the threshold of count 5. The data showcases the prominence of JavaScript and Python in contemporary hackathon projects, with both languages collectively being the primary choice for a vast majority of teams.

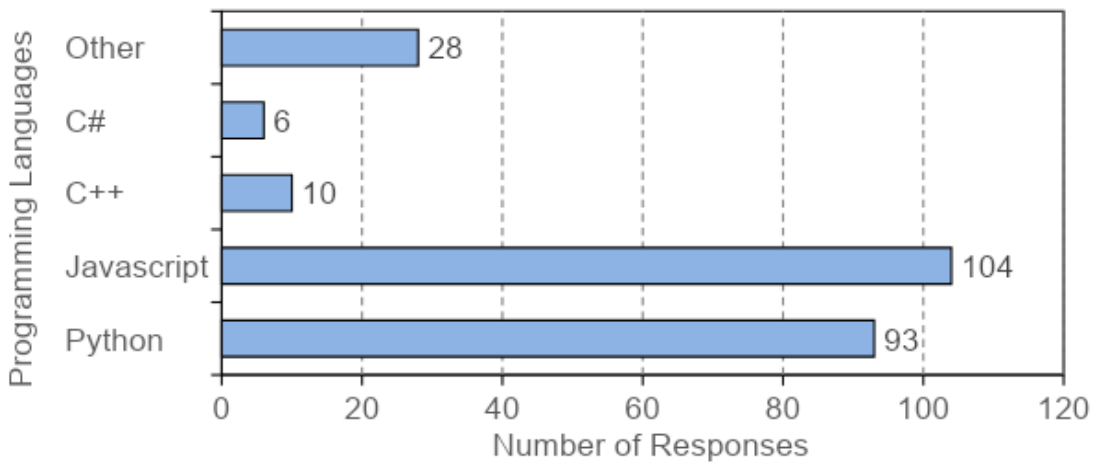

Figure 4. Programming languages used in the hackathon.

**Primary Framework and Library Selection**
Frameworks and libraries are essential in streamlining the development process, especially in hackathons where time is of the essence. They offer developers ready-made solutions and tools that can be easily integrated into projects. This section explores the most sought-after frameworks and libraries that were pivotal to the teams' efforts in the hackathon, reflecting the current trends in rapid development tools. Figure 5 illustrates the primary frameworks or libraries utilized by participants in their projects. The "Other" category, which consists of frameworks/libraries (e.g., Gradio, Next.js, Unity, Streamlit, Ruby on Rails, Vue.js.) used by fewer than 5 participants, emerged as the most popular choice, with 98 selections.

React, a widely used library for web development was selected by 71 participants, corroborating the earlier observation that web development projects were predominant in the hackathon. This choice underlines React's ability to facilitate efficient and dynamic web applications, making it a favorite in the hackathon circuit. Express, commonly paired with React for developing web applications, was opted for by 28 participants. Its popularity can be attributed to its flexibility and compatibility with Node.js, enhancing the back-end development process.

Python's influence was evident with 16 participants choosing Flask and 13 participants opting for Django. These Python-based frameworks are known for their simplicity and effectiveness in web application development, data analysis, and backend support, resonating with Python's widespread use in the hackathon. Lastly, Taipy was used by 5 participants, indicating some niche preferences or specific use cases. It's important to emphasize that teams could select multiple frameworks or libraries for their projects, and each selection was counted individually. The data not only provides a snapshot of the current tools in demand but also underscores the relationship between the type of projects and the chosen frameworks and libraries.

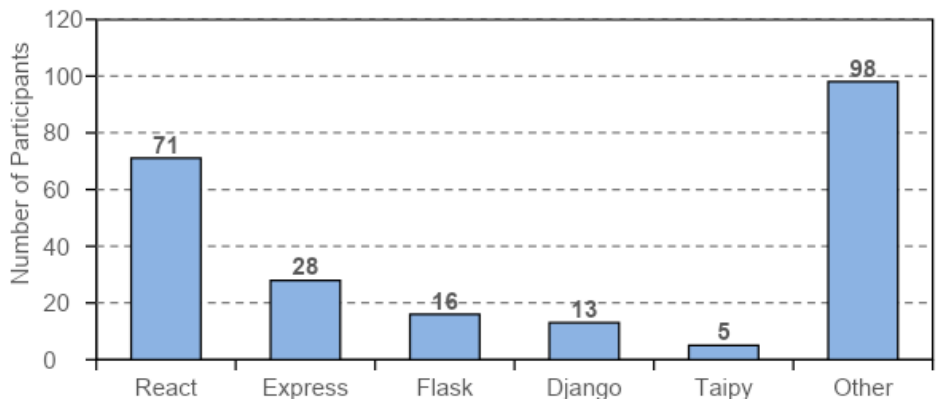

Figure 5. Frameworks and libraries used in the hackathon.

### 3.2. ChatGPT and NLP Utilization

The rise of NLP technologies, particularly tools like ChatGPT, is redefining how developers approach problems and execute solutions. This subsection takes a closer look at how hackathon participants have embraced these advancements, shedding light on the specific areas where ChatGPT and other NLP tools have been instrumental. As developers gain access to sophisticated tools like ChatGPT, understanding the extent of its integration becomes pivotal. This subsection breaks down the varying degrees to which teams have incorporated ChatGPT into their projects, revealing patterns of adoption and the potential areas where the tool has found the most traction.

Figure 6 presents an insightful depiction of how ChatGPT was integrated into team projects during the hackathon, with a scale ranging from 1 (representing 'Not at All') to 5 (indicating 'Extensively'). The data reveals a diverse spectrum of utilization levels among the participants. At the lower end of the scale, 27 participants did not incorporate ChatGPT into their projects at all, aligning with level 1. This group might have focused on areas where ChatGPT's capabilities were not pertinent or preferred other technologies.

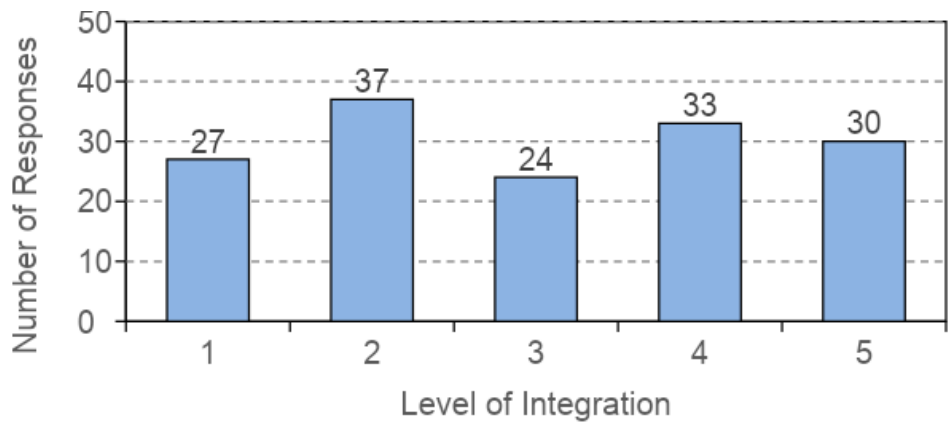

Figure 6. Usage of ChatGPT in projects based on level of integration ranging from 1 ('Not at All') to 5 (indicating 'Extensively').

A slightly larger cohort, consisting of 37 participants, minimally used ChatGPT, falling into level 2. This occasional usage suggests a selective approach, possibly integrating ChatGPT for specific tasks or exploratory purposes. In the mid-range, 24 participants indicated occasional integration of ChatGPT, categorized under level 3. This occasional use highlights a balanced approach, where ChatGPT was employed for certain aspects of the projects, reflecting its growing relevance in software development. Moving towards higher frequency, 33 participants frequently utilized ChatGPT, marking level 4.

This demonstrates a significant reliance on ChatGPT's capabilities, indicating its effectiveness in enhancing project development. At the uppermost end, 30 participants extensively employed ChatGPT, corresponding to level 5. This substantial number underscores the tool's robust applicability and its potential to significantly impact project outcomes. Notably, almost 58% of the teams used ChatGPT at a moderate level or higher (levels 3 to 5). This trend emphasizes the rising significance and adoption of such technologies in software projects. The diverse distribution reflects the varied strategies teams employed when incorporating ChatGPT into their workflow.

**Importance of NLP, including ChatGPT, in Achieving Project Goals**
The efficacy of a technology is often gauged by its perceived value in achieving specific goals. In this subsection, we explore participants' sentiments on the role of NLP technologies, especially ChatGPT, in meeting their project objectives. Through this lens, we aim to discern the overall impact and significance these tools hold in the current hackathon landscape.

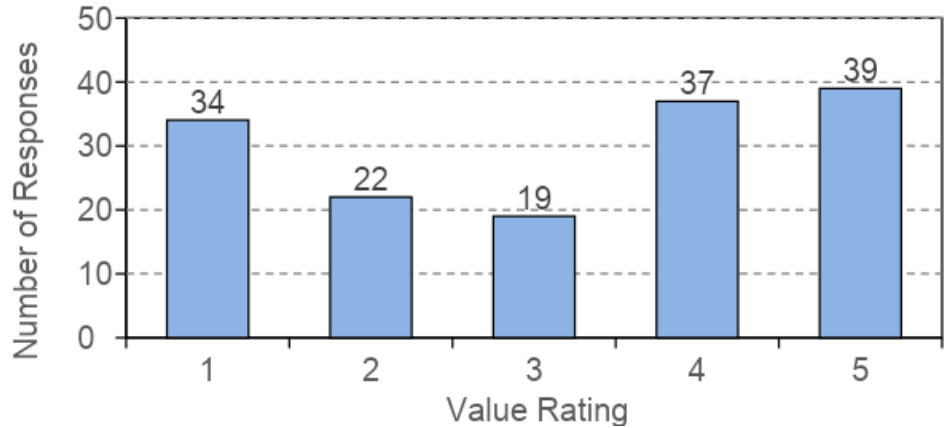
Figure 7. Perceived value of NLP technologies in achieving project goals.

Figure 7 offers a comprehensive analysis of the perceived value of NLP (Natural Language Processing) technologies, including ChatGPT, in supporting project objectives. Participants were surveyed on a scale ranging from 1 to 5, with 1 signifying "Not Valuable at All" and 5 indicating "Extremely Valuable". The data delineates intriguing patterns in participant perceptions. At the lower end of the scale, 34 participants rated NLP technologies at level 1, considering them "Not Valuable at All". This perspective might stem from specific project requirements or challenges that these technologies could not address effectively. Contrastingly, a notable segment of 39 participants considered NLP technologies to be of paramount value, granting them the highest score of 5, "Extremely Valuable".

This group recognizes the substantial impact and contribution of NLP tools in achieving their project goals. 22 participants perceived these technologies as "Slightly Valuable", assigning them a level 2 rating. This suggests a recognition of some utility, albeit limited, in their project contexts. For level 4, described as "Very Valuable", 37 participants acknowledged the significant worth of these technologies. This indicates a strong appreciation of NLP tools' capabilities in enhancing project development. 19 respondents, aligning with level 3, evaluated NLP technologies as "Moderately Valuable".

This middle-ground assessment suggests a balanced view, recognizing the usefulness of NLP tools while also acknowledging their limitations or constraints in certain applications. Additionally, it's worth noting that 63% of the teams believe NLP technologies like ChatGPT have provided moderate to significant assistance in achieving project goals. This reflects the growing importance and usefulness of these technologies in project development and underscores their increasing influence on successful project outcomes. This distribution suggests a general inclination towards the higher value of NLP technologies in project implementations, with only a subset viewing them as less impactful.

**Specific Aspects of Projects for ChatGPT or the ChatGPT API Usage**
Innovation thrives when tools are utilized in varied and unexpected ways. This section delves into the multifaceted applications of ChatGPT and its API, highlighting the diverse project facets it influenced. Figure 8 provides a detailed overview of how ChatGPT and the ChatGPT API were applied in various projects, highlighting the specific functionalities for which these technologies were utilized. The responses from the participants showcase the multifaceted applications of these tools, reflecting their adaptability and relevance in different aspects of project development. The most prominent application, as indicated by 78 participants, was "Debugging or error identification." This high number underscores ChatGPT's effectiveness in identifying and resolving coding issues, a critical aspect in software development.

Meanwhile, "Code generation or assistance" emerged as another significant use case, with 62 respondents employing ChatGPT for this purpose. This reflects the growing reliance on AI tools for enhancing coding efficiency and accuracy. For "Learning new tools or technologies," 55 participants found the use of ChatGPT and its API to be beneficial. This suggests that these technologies are increasingly being seen as valuable resources for skill development and acquiring technical knowledge. "Idea generation and brainstorming" was leveraged by 39 participants.

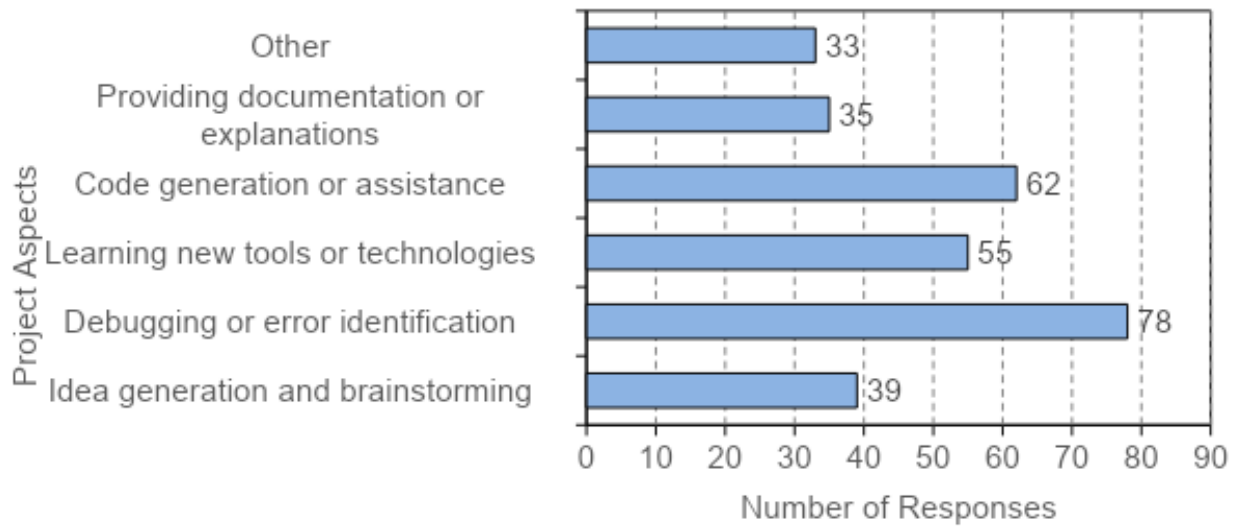

Figure 8. Usage of ChatGPT and GPT API across different project aspects.

This use case highlights the role of ChatGPT in fostering creative thinking and generating innovative solutions in the development process. Additionally, 35 individuals tapped into these tools for "Providing documentation or explanations" indicating ChatGPT's utility in simplifying complex concepts and enhancing the understanding of technical materials. Interestingly, a segment consisting of 33 respondents indicated "Other," suggesting that they did not find a fitting category or possibly did not use the technologies for the stated purposes.

This depiction underscores the versatility of ChatGPT and its API, catering to a wide spectrum of project needs. The data from Figure 8 not only underscores the versatility and widespread adoption of ChatGPT and its API across various facets of project development but also signals a significant shift in software development practices. The increasing use of AI tools like ChatGPT in tasks traditionally performed manually signifies a transformative change in the industry, reflecting the evolving role of AI in enhancing and streamlining the software development process.

### 3.3. Perception of Coding Assistance Technologies

The rapid advancements in AI and coding assistance technologies have ushered in a new era of development practices, sparking debates about their implications on traditional coding activities. This subsection navigates through the varied perceptions of hackathon participants regarding the role and impact of tools like ChatGPT and GitHub Copilot on manual coding. The rise of AI-based tools like ChatGPT and GitHub Copilot has significantly influenced software development practices. In this subsection, we examine the opinions of hackathon participants about these technologies and their effects on traditional coding. Using Figure 9 as a reference, we assess how developers feel about the balance between manual coding and automated assistance. The data reveals varied perceptions, with some seeing great potential in these tools, while others express reservations about their long-term implications.

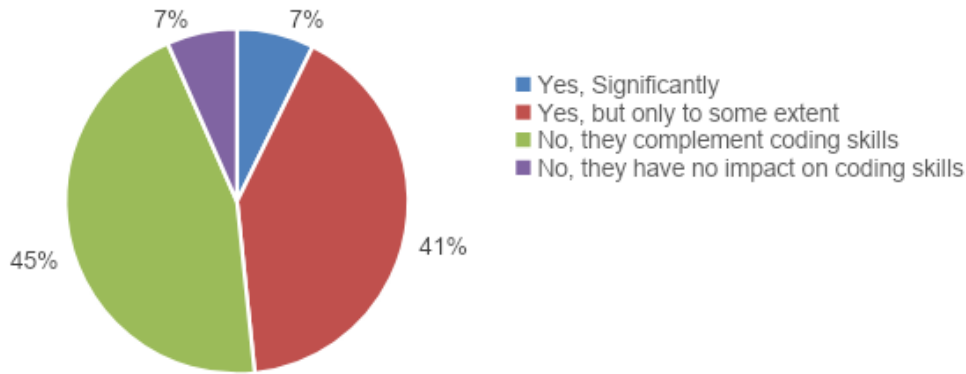

Figure 9. Impact of assistive tools on manual coding in hackathon.

Figure 9 delves into the perceptions of coding assistance tools such as ChatGPT or GitHub Copilot, particularly focusing on their impact on manual coding during hackathons. The survey, comprising 151 respondents, provides valuable insights into the evolving coding practices and the role of these technological aids in such competitive environments. Among the respondents, a notable 45% perceive these tools as helpful adjuncts that complement but do not replace the need for manual coding. This viewpoint suggests a balanced approach, where AI tools are used as a supplement to enhance coding efficiency while retaining the fundamental aspects of manual coding.

In contrast, 41% of the participants believe that these tools do reduce some of their coding workload. This perspective underscores the utility of AI tools in alleviating certain coding tasks, possibly by automating repetitive elements or providing code suggestions, thereby streamlining the development process. A smaller segment, constituting 7% of the respondents, feels that these tools significantly diminish their coding tasks. This group likely experiences a more profound impact of AI tools, perhaps relying on them for more complex coding solutions or a broader scope of assistance. Conversely, another 7% do not perceive any substantial difference made by these tools in their coding activities.

This opinion could stem from a preference for traditional coding methods or a perception that AI tools do not significantly enhance their specific coding practices. This data gives us a glimpse into how coding habits are changing and the role these tech aids play, particularly in hackathon events. The numbers suggest that the landscape of coding is shifting, with many developers starting to lean on AI tools to speed up and ease parts of the development work. This could mean more efficient and faster software creation. Yet, developers need to stay updated and adjust their skills to use these new tools properly. Most participants did find these technologies helpful, even if not groundbreaking.

### 3.4. Hackathon Communication and AI Integration

In the digital age, communication platforms like *Discord* play a crucial role in hackathon settings, fostering collaboration and knowledge sharing. This section delves into the potential synergies between AI tools and these communication platforms, exploring whether the integration of AI agents could redefine the way participants communicate during hackathons. The wave of digital transformation has not left any stone unturned, including communication platforms widely used during hackathons.

*Discord*, being a premier platform in this realm, stands at the cusp of potential AI integration. This subsection delves into the nuanced perspectives of hackathon participants on embedding AI agents into *Discord*. While many envision this as a significant leap towards more efficient communication, there are voices of caution that emphasize the potential challenges and limitations. Through an exploration of Figure 10, we aim to strike a balance between enthusiasm for AI's potential and the reservations held by some participants.

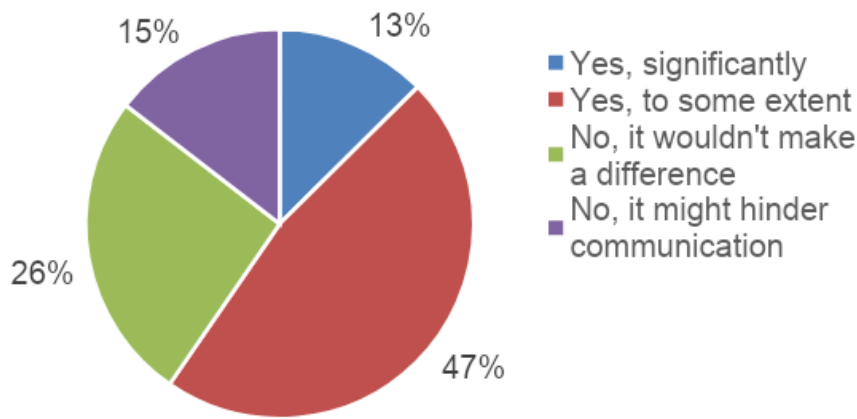

Figure 10. Perceived benefits of AI integration in Discord for hackathon communication.

Figure 10 in the study presents the responses regarding the proposition of integrating an AI agent into Discord for facilitating communication during hackathons. The feedback, collected from 151 respondents, offers insight into the participants' attitudes toward the potential use of AI in enhancing communication dynamics. A significant portion, 47% of the respondents, believe that introducing an AI agent would enhance the hackathon experience to some extent. This indicates a cautious optimism, suggesting that while there is recognition of the potential benefits of AI assistance, expectations are tempered.

Conversely, 26% of the participants feel that the inclusion of an AI agent would not make a significant difference. This group might perceive the existing communication methods as sufficient or may be skeptical about the added value of AI in this context. 14% of the respondents expressed concerns, indicating that the introduction of an AI agent might hinder communication. This cautious viewpoint highlights apprehension about potential complications or disruptions that AI could introduce into the natural flow of human interactions. Meanwhile, a smaller segment of 13% is very optimistic, stating it would improve the experience "significantly".

Analyzing the data, it's evident that while many are open to the idea of AI integration, believing it could offer some benefits, there's no overwhelming consensus on its potential impact. The majority leans positive but with moderate expectations. A combined 40% express either indifference or concerns about potential drawbacks. This mixed response suggests that while there's interest in AI-enhanced communication tools, there's also a need to address and mitigate potential challenges and ensure that the introduction of such technology aligns with user needs and doesn't disrupt the intended communication flow in hackathon settings.

**Comparison of Traditional vs. AI-Assisted Communication in Hackathon Settings**
With the proliferation of AI-powered tools offering real-time support and guidance, there's a growing interest in their efficacy compared to traditional human moderators. This subsection offers insights into participants' preferences when it comes to question-answering support during hackathons, emphasizing the balance between human touch and AI assistance.

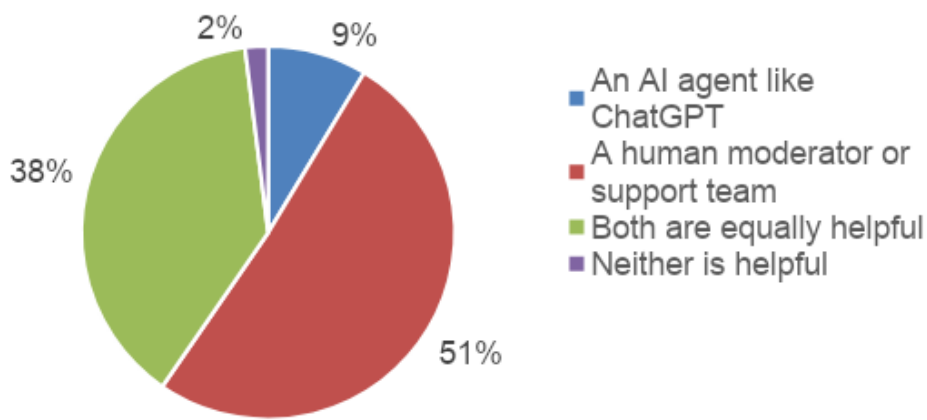

Figure 11. Preferences for question-answering support during hackathons on Discord.

Figure 11 provides an insightful analysis of the preferred resources for addressing inquiries during hackathons, particularly on platforms like Discord. The survey, encompassing 151 participants, sheds light on the balance between human and AI-based support preferences. A significant 51% of respondents expressed a preference for "A human moderator or support team" as their primary resource for assistance. This majority indicates a strong inclination towards human interaction and expertise, reflecting the value placed on personal and context-specific guidance in resolving queries.

Interestingly, a notable 38% of participants perceive "Both (a human moderator and an AI agent like ChatGPT)" as equally helpful. This response points to a growing recognition of the complementary strengths of human and AI resources. It suggests an appreciation for the efficiency and scalability of AI tools like ChatGPT, coupled with the nuanced understanding and adaptability provided by human moderators. Only 9% prefer solely relying on "An AI agent like ChatGPT". A very small fraction, at 2%, think that "Neither is helpful".

This data demonstrates that while human support is still the top preference for many during hackathons, there's a substantial group that sees value in a combined approach of both human and AI resources. The relatively low percentage preferring only AI suggests that while AI tools like ChatGPT have their place, they might not yet be seen as full replacements for human interaction in such settings. Thus, for optimal user experience during hackathons, organizers might consider a hybrid support model, blending human expertise with AI capabilities, to cater to the broad preferences of participants.

**Impact of AI on the Traditional Hackathon Experience**

Hackathons have long been seen as the epitome of innovation, collaboration, and hands-on coding. As AI-driven code-generation tools make their presence felt, there's a pressing need to understand their influence on the quintessential hackathon vibe. This section captures the sentiments of participants, gauging the extent to which technologies like ChatGPT are perceived to be reshaping the hackathon narrative.

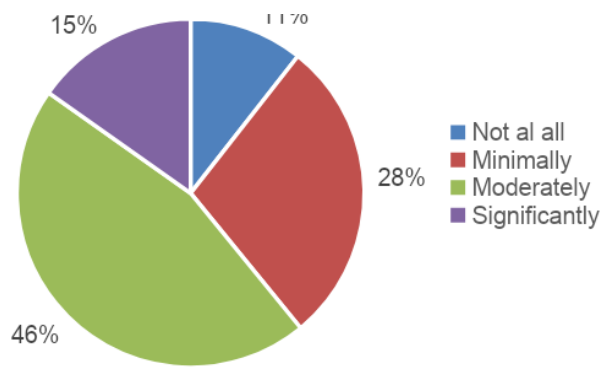

Figure 12. Perceived influence of code-generating AI on traditional hackathon dynamics.

Figure 12 examines the viewpoints of hackathon participants regarding the impact of code-generating technologies like ChatGPT on the conventional hackathon experience. The survey, encompassing 151 participants, offers a nuanced understanding of how these AI tools are perceived in shaping the dynamics of hackathons. A plurality of 46% of the respondents believe that these tools bring about a "Moderate" change to the traditional hackathon format. This perspective suggests that while these technologies are influential, their impact is more evolutionary than revolutionary, enhancing certain aspects of the hackathon experience without completely overhauling it.

Next, 28% of participants perceive the influence of these tools as "Minimal." This group likely views AI technologies as useful additions that subtly integrate into the existing framework of hackathons, without dramatically altering their fundamental workflow. 15% of the respondents see a "Significant" impact from these technologies. This group might recognize the potential of AI tools to greatly enhance productivity, creativity, or efficiency in hackathons, possibly seeing them as key drivers of future changes in the format and execution of these events. Conversely, 11% feel that these technologies do "Not at all" change the classic hackathon experience.

This opinion could stem from a view that the essence of hackathons – collaboration, innovation, and rapid development – remains unchanged despite the introduction of new technological tools. The information paints a picture where a considerable chunk of respondents sees these AI technologies as having some influence on hackathons, yet not overwhelmingly so. The dominant sentiment is that while these tools bring changes to the table, they don't radically redefine the entire hackathon experience. The blend of responses indicates an ongoing transition in the tech community's perception, where AI's role in hackathons is acknowledged but not seen as a complete game-changer. Organizers and participants alike should be aware of this sentiment when considering the adoption and implementation of such technologies in future events.

## 4. Discussion

In this section, we examine the implications of our findings in relation to the integration of AI technologies, such as ChatGPT, into hackathons and educational settings. We consider both the potential benefits, including enhanced productivity and inclusivity, as well as challenges like ethical concerns and impacts on collaboration. By situating our results within the context of existing research, we aim to provide a balanced perspective on the role of AI in shaping learning experiences and offer recommendations for its responsible and effective use in future hackathons.

### 4.1. The Role of AI in Shaping Hackathon Experiences

The inclusion of AI tools like ChatGPT in hackathons can indeed transform how these events operate, offering both opportunities and challenges. Our results showed that participants widely used AI tools to boost efficiency, especially in speeding up coding tasks, brainstorming solutions, and generating ideas. This aligns with prior research, which noted that AI could significantly accelerate project completion (Lau & Guo, 2023). The ability of AI to offer instant feedback and sophisticated problem-solving approaches (Liu et al., 2024) was highlighted by participants as a key advantage, improving the overall quality of projects.

However, our findings suggest that AI's role is not without complications. While AI facilitates quick solutions, it may affect team dynamics. Some participants reported a tendency to rely heavily on AI-generated solutions, which in some cases led to reduced collaboration and fewer discussions within teams. This finding resonates with concerns raised in the literature about the potential for AI to bypass critical thinking and manual coding skills (Zastudil et al., 2023). In particular, participants with less experience admitted to over-relying on AI tools, reducing their engagement in the problem-solving process. This aligns with Hou et al.'s (2024) observation that while AI help-seeking is becoming prevalent, it risks replacing traditional, skill-developing resources, especially for beginners.

The skills cultivated during AI-driven hackathons, particularly in navigating AI tools like ChatGPT, are likely to have far-reaching implications for the future of work. As companies increasingly integrate AI into their workflows, workers who are proficient in these tools may experience greater job security and mobility, while those without access or training in AI may face barriers to entry in an increasingly automated labor market. This shift could exacerbate existing inequalities in access to technology, further entrenching divides between experienced developers and those new to the field. Furthermore, the reliance on AI to complete tasks typically associated with human creativity and problem-solving raises critical questions about the value placed on manual coding skills and the implications for long-term employment prospects in the tech industry.

Conversely, AI tools were also seen to "level the playing field" by helping beginners close the knowledge gap with more experienced developers (Prather et al., 2023). This inclusivity is a key benefit, as it allows participants from diverse backgrounds to engage more fully in hackathons. Yet, the overuse of these tools could risk compromising the spirit of collaboration central to hackathons, echoing concerns about ethics and educational integrity raised by Kooli (2023).

Additionally, ethical issues, such as plagiarism and the fairness of using AI in competition, were raised by both participants and judges. Our results showed that some teams leveraged AI tools without fully understanding their outputs, raising concerns about attribution and originality. This reinforces previous studies' warnings regarding the balance between using AI as a supportive tool versus letting it dominate the creative process (Chan, 2023). Judges in particular expressed difficulty in evaluating projects that heavily relied on AI, pointing out the challenges in distinguishing AI-generated work from the team's genuine efforts. This highlights the need for transparent and robust guidelines for AI usage in hackathons, as suggested by Krishnamurthy et al. (2018). Over-dependence on these tools might reduce collaboration and teamwork among participants, potentially undermining the collaborative spirit of hackathons. Evaluating projects heavily reliant on AI can also be challenging for judges, who must balance AI assistance and individual effort. Finally, issues related to accessibility, technical glitches, and data privacy and security must also be carefully managed. While incorporating AI into hackathons offers the potential to enhance productivity, learning, and collaboration, it also poses challenges related to ethics, over-reliance on these tools, and the need for appropriate evaluation methods. Organizers and participants must find a balance between harnessing the benefits of AI and preserving the essential elements of traditional hackathon experiences.

## 4.2. Implications for Education

The changes happening in artificial intelligence reflect larger shifts in technology and engineering education, including the integration of advanced tools like ChatGPT into education and professional training (Pursnani et al., 2023), and the rise of Virtual Teaching Assistants (VirtualTAs), powered by AI tools (Grassini, 2023). These offer an innovative way to bridge the gap between traditional classroom learning and the real-world challenges of hackathons. These digital mentors are readily available, providing instant and personalized support to participants, and adapting to their unique learning styles and needs (Sermet and Demir, 2021). This scalability is especially beneficial in the fast-paced hackathon environment, where many participants often need simultaneous guidance.

VirtualTAs offer benefits like immediate feedback, help with complex problem-solving, and improved collaboration among team members, encouraging a hands-on learning approach. Furthermore, AI technologies play a crucial role in learning analytics, a field that leverages data collection and analysis to enhance the educational process. By integrating learning analytics, these technologies can track and analyze students' performance, engagement, and learning patterns (Sajja et al., 2023c). This data-driven approach enables educators and AI tools to tailor their teaching methodologies and resources to the specific needs of each student, thus optimizing the learning experience. Advanced analytics can predict student performance, identify at-risk students early, and provide interventions tailored to individual learning curves.

This aspect of AI not only personalizes education but also improves the overall effectiveness and efficiency of the learning process. The use of these technologies makes tech education more globally accessible, aligning with the industry's evolving demands for smart tools and personalized experiences. However, it's essential to consider the feedback from hackathon participants. Interestingly, despite the evident advantages of VirtualTAs, 51% of participants still expressed a preference for in-person assistance. This suggests potential reservations or concerns about fully relying on AI-driven platforms for education, emphasizing the continued importance of human interaction in learning environments.

Beyond hackathons, the use of AI in education can also reduce the workload on instructors and teaching assistants. By using virtual teaching assistants, educational institutions can handle logistical questions about courses, creating a more efficient learning experience (Sajja et al., 2023a). This technology provides easy access to information and offers personalized learning support, including generating flashcards and quizzes on various course topics, giving students flexibility in their studies (Sajja et al., 2023b). The broader transformation in customizable tools driven by AI creates enhanced avenues for research communication (Sermet and Demir, 2019) and educational settings like hackathons and traditional classrooms.

## 4.3. Recommendations

Integrating AI technologies into hackathons should be guided by a set of best practices to ensure a productive and ethical environment. Organizers must establish clear objectives and guidelines for AI usage, defining specific areas where these technologies can be beneficial while encouraging participants to focus on creative problem-solving. Pre-hackathon workshops and educational sessions should be conducted to familiarize participants with AI tools, fostering a foundational understanding of technological scope. Ethical considerations should be at the forefront, emphasizing responsible usage, plagiarism avoidance, and proper attribution.

Providing mentorship and support from experienced professionals can help participants strike a balance between required assistance and critical thinking coding. Inclusivity can be achieved by assessing the diverse skill levels among participants, with different challenges tailored for beginners and advanced hackers, promoting collaboration within teams. Real-world problem-solving should be a focus, aligning hackathon challenges with issues AI can help address, and encouraging participants to consider broader implications. Transparent judging criteria and feedback mechanisms should be established, and post-hackathon resources should be provided to facilitate continued skill development.

## 5. Conclusion

This study highlighted crucial insights into the integration of AI and NLP technologies, specifically ChatGPT, within the context of hackathon events, and it prompts thoughtful consideration of their implications for the evolving landscape of educational hands-on experiences. Based on the findings from our research questions, we can deduce several key insights. NLP technologies, notably exemplified by ChatGPT, have carved out an increasingly pivotal role in hackathon projects. Their ascent aligns with the broader recognition of their potential to boost efficiency, stimulate creativity, and enhance problem-solving capacities among participants.

In the context of hackathons, ChatGPT's most notable application is evident in the realms of code debugging or error identification. This tool serves as a robust catalyst, enabling participants to swiftly pinpoint and rectify errors, navigate through coding challenges, and refine their project implementations. However, as we progress into an era where tools like ChatGPT and GitHub Copilot emphasize automated coding, we observe a nuanced shift in the valuation of manual coding skills. While the allure of these tools is grounded in their promise of amplified efficiency, there's a looming risk. An over-reliance on automation might inadvertently undermine the cultivation of critical thinking and foundational coding skills.

These competencies remain at the heart of sustainable growth in the realm of software development. In tandem with these developments, the potential of AI-driven communication tools is beginning to be realized. Platforms like Discord, when synergized with AI-driven tools, present a tantalizing prospect. They could redefine the dynamics of collaborative communication during hackathons, providing participants with rapid, context-aware responses that bolster collaborative problem-solving. However, it's essential to consider the feedback from hackathon participants. Interestingly, despite the evident advantages of VirtualTAs, 51% of participants still expressed a preference for in-person assistance.

This suggests potential reservations or concerns about fully relying on AI-driven platforms for education, emphasizing the continued importance of human interaction in learning environments. In the future, the increased usage of AI tools within hackathons is poised to continue influencing the landscape of these events. Anticipated trends include even deeper integration of AI-driven technologies throughout various stages of project development. However, as the role of AI intensifies, it becomes imperative for hackathon organizers and participants alike to grapple with the ethical considerations surrounding its use, setting up the correct balance between required assistance and the cultivation of manual coding skills, and develop fair evaluation mechanisms to safeguard the integrity of these events.

As AI continues to shape hackathons and education, the skills gained in these settings will likely impact the broader workforce. AI is increasingly automating tasks and workflows, and workers skilled in AI-driven tools will be essential as industries adopt these technologies. However, this raises important concerns about labor market inequality, skill gaps, and the future of work, where human creativity and problem-solving may become mediated by AI.

The use of AI tools in hackathons will keep growing, further integrating into project development processes. Organizers must balance AI assistance with the cultivation of manual coding skills and develop fair evaluation methods to maintain the integrity of these events. In a broader sense, hackathons are evolving with technological advancements, emphasizing the need for professionals proficient in AI. These events continue to bridge the gap between classroom learning and real-world applications, fostering the skills needed to meet the industry's evolving demands.

**Availability of Data and Materials**
All data that is produced and analyzed in the manuscript is readily available and presented in the manuscript.

**Competing Interests**
The authors declare that they have no competing interests.

**Declaration of generative AI and AI-assisted technologies in the writing process**
During the preparation of this manuscript, the authors used ChatGPT, based on the GPT-4 model, to improve the flow of the text, correct grammatical errors, and enhance the clarity of the writing. The language model was not used to generate content, citations, or verify facts. After using this tool, the authors thoroughly reviewed and edited the content to ensure accuracy, validity, and originality, and take full responsibility for the final version of the manuscript.